\documentclass[preprint,twocolumn,12pt,5p]{elsarticle} %review,12pt,1p %preprint,twocolumn,12pt,5p
\usepackage[utf8]{inputenc}
\usepackage{graphicx}
    \graphicspath{ {Figures/} }
\usepackage{float}
\usepackage{caption}                            
\usepackage{amsmath, amsthm, amssymb}
\usepackage{mathtools}                        
\usepackage[version=4]{mhchem}
\usepackage[linktoc=all,pdfencoding=auto, psdextra]{hyperref}
\hypersetup{pdfauthor=author}
\usepackage{lineno}
\usepackage{orcidlink}
\usepackage{natbib}

\journal{Nuclear Instruments and Methods:A}

\date{\today}

\begin{document}

    \begin{frontmatter}
    \title{High-accuracy liquid-sample \(\beta\)-NMR setup at ISOLDE}
    %Authors
\author[CERN,UniGe]{J. Croese\corref{cor1}\orcidlink{0000-0002-0660-4409}}
    \ead{jared.croese@cern.ch}
    \cortext[cor1]{Corresponding authors}
\author[Poznan]{M. Baranowski\orcidlink{0000-0002-8584-7029}}
\author[Manchester]{M. L. Bissell\orcidlink{0000-0002-0144-5015}}
\author[CERN,Leipzig]{K. M.  Dziubinska-K{\"u}hn\orcidlink{0000-0002-4526-841X}}
\author[Leuven]{W. Gins\orcidlink{0000-0002-2353-7455}\fnref{fn1}}
    \fntext[fn1]{Current address: University of Jyv{\"a}skyl{\"a}, Jyv{\"a}skyl{\"a}, Finland}
\author[CERN,York]{R. D. Harding\orcidlink{0000-0002-6312-783X}\fnref{fn2}}
    \fntext[fn2]{Current address: Nottingham University Hospitals Trust, Nottingham, United Kingdom}
\author[Maastricht]{R. B. Jolivet\orcidlink{0000-0002-5167-0851}}
\author[Leuven]{A. Kanellakopoulos\orcidlink{0000-0002-6096-6304}\fnref{fn3}}
    \fntext[fn3]{Current address: HES-SO, Geneva, Switzerland}
\author[UniGe]{B. Karg\orcidlink{0000-0003-2648-3509}}
\author[CERN,UniGe]{K. Kulesz\orcidlink{0000-0002-7108-0939}}
\author[Tennessee]{M. Madurga Flores\orcidlink{0000-0002-8177-4328}}
\author[CERN,Leuven]{G. Neyens\orcidlink{0000-0001-8613-1455}}
\author[CERN]{S. Pallada\orcidlink{0000-0002-4309-6458}\fnref{fn3}}
\author[Poznan]{R. Pietrzyk}
\author[Warsaw]{M. Pomorski\fnref{fn4}}
    \fntext[fn4]{Current address: Univ. Bordeaux, CNRS, CENBG, UMR 5797, F-33170 Gradignan, France}
\author[CERN,Oldenburg]{P. Wagenknecht\fnref{fn5}}
    \fntext[fn5]{Current address: University of Tennessee, Knoxville, USA}
\author[Czech]{D. Zakoucky}
\author[CERN]{M. Kowalska\orcidlink{0000-0002-2170-1717}\corref{cor1}}
    \ead{magdalena.kowalska@cern.ch}

%Affiliations
\address[CERN]{CERN, Geneva, Switzerland}
\address[UniGe]{University of Geneva, Geneva, Switzerland}
\address[Poznan]{Adam Mickiewicz University, Poznan, Poland}
\address[Manchester]{University of Manchester, Manchester, United Kingdom}
\address[Leipzig]{Leipzig University, Leipzig, Germany}
\address[Leuven]{KU Leuven, Leuven, Belgium}
\address[York]{University of York, York, United Kingdom}
\address[Maastricht]{Maastricht University, Maastricht, The Netherlands}
\address[Oldenburg]{Oldenburg University, Oldenburg, Germany}
\address[Tennessee]{University of Tennessee, Knoxville, USA}
\address[Warsaw]{University of Warsaw, Warsaw, Poland}
\address[Czech]{Nuclear Physics Institute of the Czech Academy of Sciences , Rez, Czech Republic}

    \begin{abstract}
Recently there has been an increased interest to apply the sensitive $\beta$-decay asymmetry detected nuclear magnetic resonance ($\beta$-NMR) technique to biological studies.
A liquid-sample $\beta$-NMR setup was build at ISOLDE to allow such investigations and to use the resolution gain of liquid-state NMR in nuclear physics.
As part of this setup a magnetic field locking system, a set of printed circuit board shimming coils, a sample exchange system, a set of compact $\beta$-detectors and a custom experimental vacuum chamber were developed.
The main magnetic field was stabilized down to the ppm level by the locking system while allowing the direct determination of the absolute magnetic field.
The homogeneity of the magnetic field was improved to $\leq$~5~ppm over the sample volume by the shimming coils.
Time spent on changing samples was reduced by a factor of five by the liquid sample exchange system.
During experiments it was possible to continuously observe the liquid sample thanks to the custom chamber and compact $\beta$-detectors.
The absolute field determination allows for a novel way to reference $\beta$-NMR measurements, removing the need for time consuming reference measurements.
The improved accuracy and resolution resulting from these innovations allows the study of the distribution of nuclear magnetization and (bio)chemicals using high-accuracy liquid-sample $\beta$-NMR. 
\end{abstract}
    \end{frontmatter}
    
%\linenumbers
\section{Introduction}\label{par:introduction}
The emission of $\beta$-radiation from an ensemble of oriented nuclei can exhibit a high degree of asymmetry.
This was first observed by Wu \textit{et al.}, with a low temperature spin-oriented sample of \ce{^{60}Co} nuclei proving that the weak force does not conserve parity\cite{Wu1957}.
Due to the magnitude of the observed asymmetry compared to other types of anisotropically emitted radiation, this effect is a promising probe for several nuclear effects\cite{PostmaStone1986}.
In particular, it can be used as a very sensitive means of detecting nuclear magnetic resonance (NMR) signals, which can be up to ten orders of magnitude more sensitive when compared to conventional NMR on stable isotopes\cite{Gottberg2014} as it allows one to measure signals from as few as 10$^7$ nuclei.

The $\beta$-asymmetry that is observed in $\beta$-decay asymmetry detected NMR ($\beta$-NMR) experiments, scales linearly with the nuclear polarisation\cite{PostmaStone1986}, i.e. the amount of first order nuclear orientation within an ensemble.
Therefore, the larger the polarisation is, the better the signal to noise ratio (SNR) is.
The high level of polarisation of the ensemble of nuclei is one of the reasons for the high sensitivity of the technique, the other being the detection by $\beta$-radiation.

The Larmor precession frequency of the nucleus and the spin-lattice relaxation time are directly affected by the electromagnetic (em) environment it experiences.
Therefore, the implanted nuclei can be used as probes of these local environments in the host material by measuring the resonance (Larmor) frequencies and relaxation times in $\beta$-NMR experiments.
To do so, nuclei that have a non-negligible beta-decay asymmetry parameter, that can be spin-polarised and have a short lifetime are required. Examples of such nuclei are \ce{^{8,11}Li}\cite{Neugart2008}, \ce{^{31}Mg}\cite{Neyens2005MeasurementState}, \ce{^{26,27,28}Na}\cite{Harding2020MagneticBiology, Kowalska2017}.
Other experimental requirements are: a stable and homogeneous magnetic field (either externally applied or internal to the sample), a radio-frequency (rf) source, a set of beta-detectors and the sample to be studied.

Radioactive ion beam (RIB) facilities, of either the in-flight or ISOL-type\cite{Blumenfeld2013}, provide many nuclei that are suitable for $\beta$-NMR.
Our experiment is located at ISOLDE, which falls into the ISOL-type category\cite{Borge2016, Borge2017}.
This facility uses a 1.4~GeV proton beam as a light projectile impinging on heavy target materials, containing e.g. uranium or lead\cite{Gottberg2016}.
The diverse mixture of simultaneously produced isotopes is selectively ionised and consecutively accelerated, mass-separated, and electrostatically guided to different experimental setups\cite{Catherall2017}. 

The resulting low-energy ion beam does not exhibit any nuclear polarisation.
Laser optical pumping, the method used at our polarisation beamline at ISOLDE \cite{Kowalska2017, Gins2019}, is a very efficient method to produce a sample of highly oriented short lived isotopes\cite{Kowalska2017}.
This is however dependent on an available closed excitation-deexcitation loop which is generally available for alkali and alkaline-earth metals. Moreover, when multiple-frequency pumping is used other elements (e.g Argon)\cite{Gins2019} can be efficiently polarized using this technique as well.

Using an accelerating/decelerating electrode the ion beam is tuned into resonance with the laser light to allow for the optical pumping process to happen, next (in the case of alkali metals) the beam passes through a charge exchange cell which neutralises the beam.
A space of 2~m is left for the atoms to interact with the laser light, after which the beam goes through a transition magnetic field which allows the spins to adiabatically rotate into the direction of the main magnetic field (B$_{0}$)\cite{Gins2019,Kowalska2017}.
Next the beam is implanted into the sample material from where the $\beta$-radiation is emitted and observed by a set of detectors located with and against the direction of B$_{0}$. After each ion bunch, the $\beta$-count asymmetry between the two detectors is recorded.
To record a resonance spectrum the frequency of the applied rf-field is modified at each bunch.
To extract from the measured Larmor frequency either the nuclear g-factor or the magnetic field felt by the nuclei, one usually performs a reference measurement.
A reference measurement can be performed by implanting another isotope of the same element in the same sample, yielding then a ratio of g-factors.
Alternatively, the isotope of interest can be implanted in another sample, yielding then a ratio of the magnetic fields felt by the isotope (and thus sensitivity to e.g. a chemical shift or a Knight shift).

$\beta$-NMR measurements have been applied in nuclear structure studies and in material science studies on solid-state samples compatible with the high vacuum environment present at RIB facilities\cite{kowalska2021}.
In nuclear structure studies, they can be used to determine magnetic dipole and electric quadrupole moments of short-lived nuclei \cite{Neyens2005MeasurementState,Arnold1987,Geithner1999MeasurementNucleus11Be,Keim2000}.
In material science, they serve as a sensitive local probe of the internal and interface em-fields\cite{MacFarlane2015}.

Recently, there has been increased interest in extending the scope of applications of the technique to (bio-)chemical studies, which typically make use of liquid samples in which small chemical shifts of the NMR signal are measured\cite{Gottberg2014,Jansco2017}.
This goal poses several design and engineering challenges: first, the need to combine the liquid environment with high vacuum and second, the requirement of a part-per-million (ppm) homogeneous and stable magnetic field.
Additionally, several practical aspects that improve the operation and throughput of the setup are also important to consider.
Such as minimising the footprint of the $\beta$-detectors and the time spent on changing liquid samples.

Here, we report on the developments that have led to the high-accuracy liquid-sample $\beta$-NMR experiments using our experimental setup at ISOLDE\cite{Gins2019,Kowalska2017}.
The upgrades described below have enabled us to determine with unprecedented accuracy the nuclear magnetic dipole moment of the short-lived \ce{^{26}Na} isotope\cite{Harding2020MagneticBiology}.
\section{Methods}

\subsection{Field locking}\label{par:Locking}
B$_{0}$ (1.2~T) was supplied by a Bruker (Billerica MA) BE25 electromagnet corresponding to a \ce{^1}{H} resonance frequency around 52~MHz.
This is generally considered low-field for modern day NMR spectrometers\cite{Blmich2019,Minkler2020}.  
Due to the crowded environment of the ISOLDE experimental hall, which contains lots of devices that generate (fluctuating) stray fields, fluctuations of B$_{0}$ are to be expected.
However, a stable B$_{0}$ is essential for $\beta$-NMR experiments, especially those performed in liquid samples, where chemical shifts of a few ppm are often measured\cite{InorgChemEncycl2011}.
These shifts can easily become unidentifiable by peak broadening due to poor magnetic field homogeneity and/or stability.
Therefore a magnetic field stability and homogeneity on the level of a few ppm or less is required.
Systematic B$_{0}$ measurements showed that besides rapid fluctuations of low amplitude a significant drift could be observed.

In order to reduce these field changes and to stabilize the magnetic field, a feedback system was implemented.
As part of this system the magnetic field was measured using the pulsed NMR method on \ce{^{1}H} inside \ce{H_{2}O} enclosed in a 3~mm (outer) diameter vacuum-tight vial.
The applied pulse length was tuned to cause a $\pi/4$ rotation of the magnetization vector.
The free induction decay (FID) of \ce{^{1}H} nuclei in water was measured using a transmitter/receiver coil tightly wound around the outside of the vial.
This signal was acquired by a PS2204a PicoScope (Pico Technology, Cambridgeshire, UK).
A LabVIEW (National Instruments, Austin TX) \cite{Elliott2007NationalMeasurement} program was used to compute a zero-filled Fast Fourier Transform (FFT) of the acquired signal to obtain the \ce{^{1}H} Larmor frequency, which is proportional to the magnitude of the magnetic field.
This frequency was in turn compared to the set point using the LabVIEW built-in PID control function.
The output of the PID function was used to determine the number of steps required by a stepper-motor driving a 10~k$\Omega$ variable resistor, connected in parallel to the magnet.
The variable resistor was in series with two 1~k$\Omega$ resistors, to avoid short-circuiting the power supply.
When coupled to the magnet's power supply (providing 38~A) the stepper-motor afforded an adjustment range of B$_{0}$ of 350~ppm.

The \ce{^{1}H} frequency measurement was repeated every 200 ms.
A faster repetition rate was not feasible due to the long spin lattice relaxation times ($T_{1}$) of \ce{^{1}H} in water, and the subdued signal strength of the FID in a low magnetic field.

\subsection{Absolute-field measurement}\label{par:absolute-field}
The measured \ce{^{1}H} frequency also allowed the determination of the absolute value of the external magnetic field $B_0$ at the position of the \ce{^{1}H} magnetometer, according to the following equation:

\begin{equation}
    B_0 = \frac{\nu_L h I}{\mu (1-\sigma)}
    \label{Eqn:extmagfield}
\end{equation}

\noindent Here, $\nu_L$ is the measured Larmor frequency, $h$ Planck's constant, $I$ the proton spin, $\sigma$=25.71(3)~ppm is the absolute shielding of \ce{^{1}H} in water at 300~K\cite{Garbacz2012,Makulski2018}, $\mu$ is the magnetic moment of the proton 2.79284734462(82) $\mu_N$ \cite{Schneider2017}.
To include the effect of the magnetic susceptibility of the probe, see Eq. (2) in ref. \cite{Harding2020MagneticBiology}. 

To determine the magnetic field in the center of the magnet, where the sample is located, the field difference between the magnetometer and the sample positions has to be taken into account.
We have done so by measuring the field with the \ce{^{1}H} magnetometer at both locations, and the relative difference amounted to 20(3)~ppm.
The biggest source of uncertainty in the determination of the absolute magnetic field at the location of the sample was due to the uncertainty in the probe's position.
\subsection{Magnetic field shimming}
\begin{figure}
    \centering
    \includegraphics[height=\columnwidth]{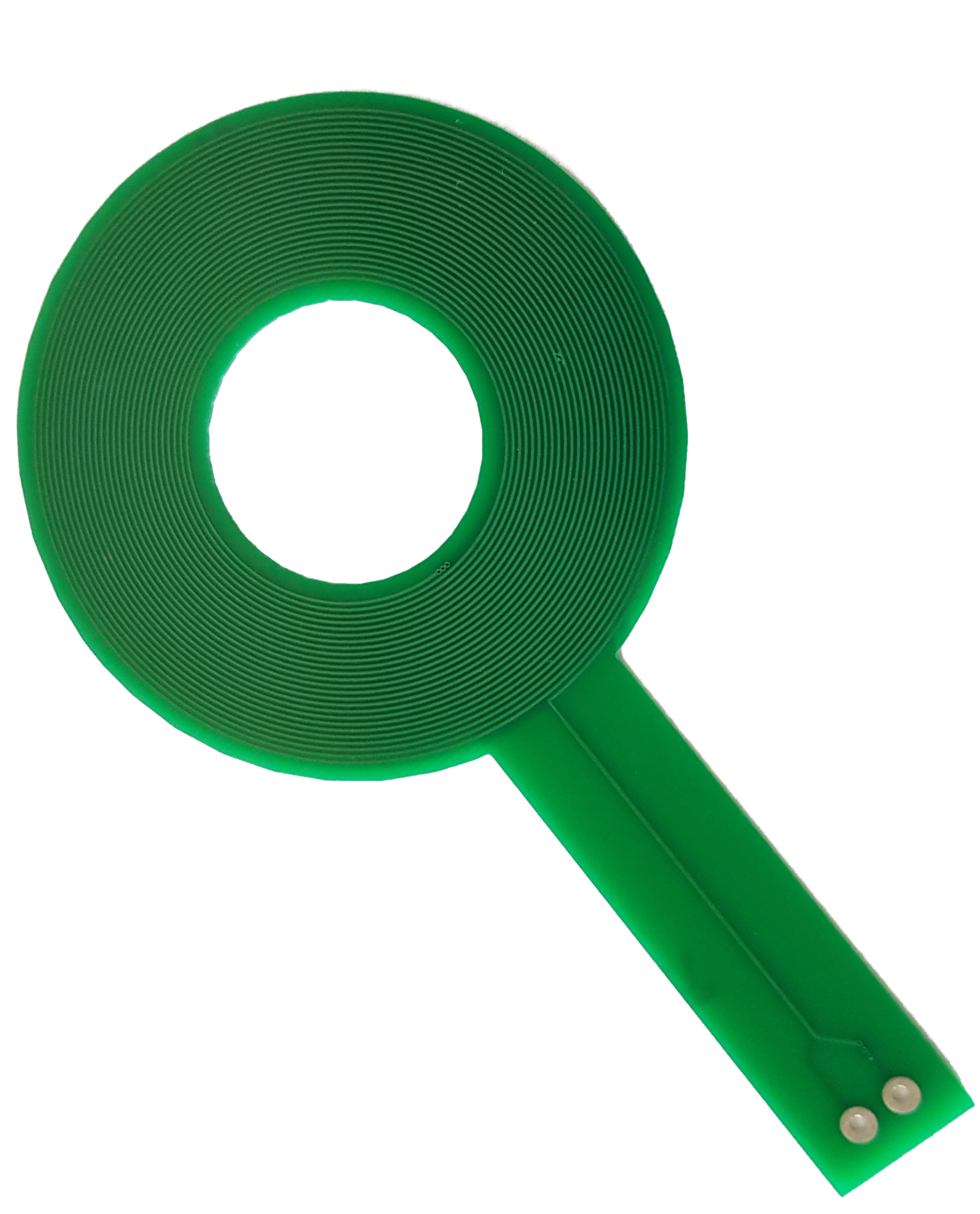}
    \caption{Detailed picture of a double layered printed circuit board shim coil.}
    \label{fig:ShimCoil}
\end{figure}

To address the 2$^{nd}$ order spatial fluctuations, or inhomogeneities, of the magnetic field in the center of the magnet, two shimming coils were designed.
They were installed concentric and in contact with the magnet poles.
For the design of the coils, the patent of G{\"u}nthard \emph{et al.} was used \cite{Gunthard1960}, which states that a pair of shimming coils placed directly against the poles of a magnet with a mean radius (A) of 0.43 times the distance between the poles (G) can effectively correct for 2$^{nd}$ order inhomogeneities, without introducing any inhomogeneities of the 4$^{th}$ order.
G{\"u}nthard \emph{et al.}\cite{Gunthard1960} also describe that the required current (I) can be calculated with:

\begin{equation}\label{eq:ShimCur}
    I = \frac{2 B_2 A^3}{N \mu_0 1.36}
\end{equation}

\noindent where $\mu_0$ is the magnetic permeability of air, N the number of coil windings and $B_2$ the second order inhomogeneous contribution to the field.
For ease of use and precision, the coils were manufactured as a double-layered printed circuit board (PCB). Figure \ref{fig:ShimCoil} shows a picture of one of the shimming coils.

\subsection{\(\beta\)-detectors}
\begin{figure}
    \centering
    \includegraphics[width=\columnwidth]{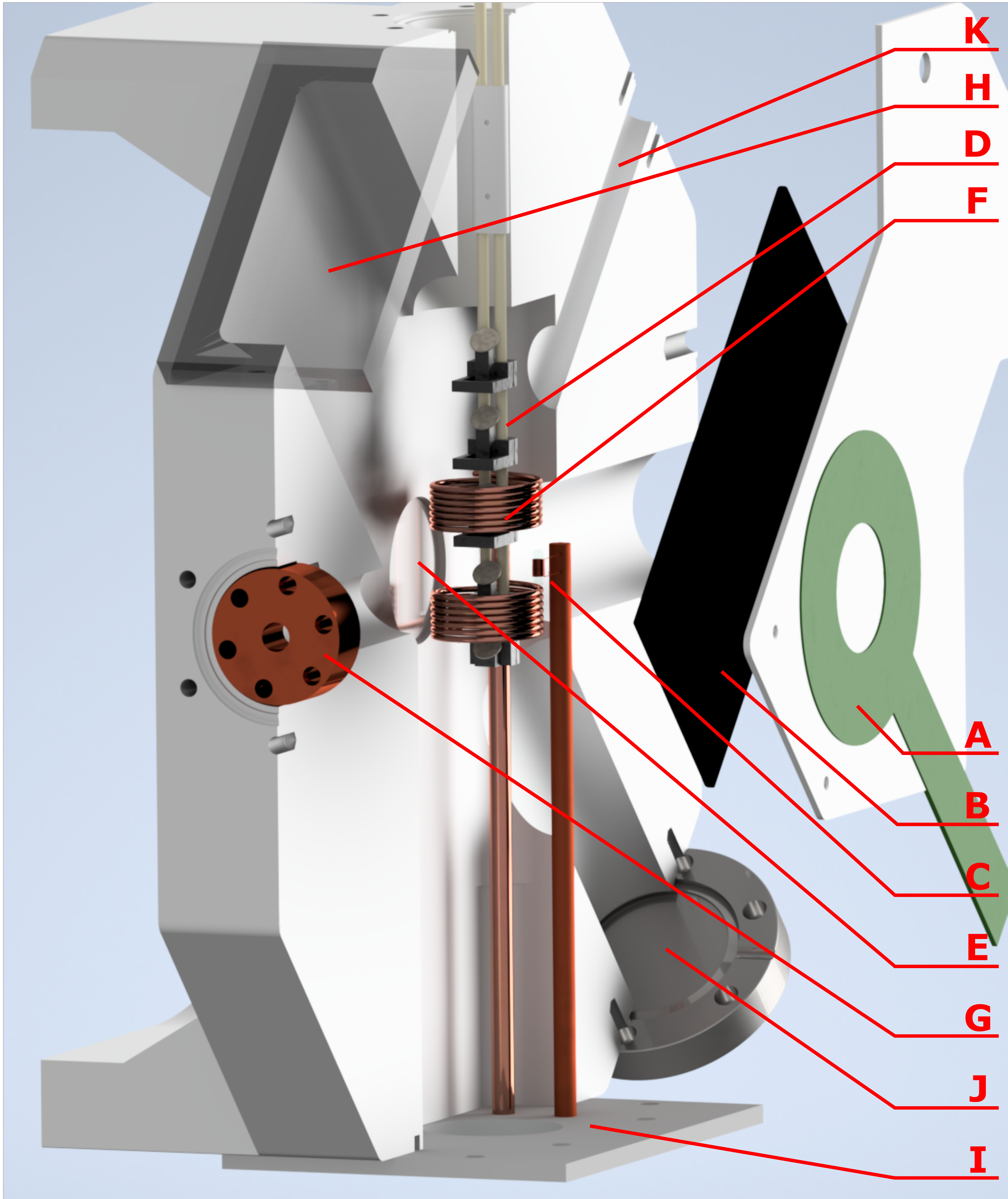}
    \caption{Exploded view of the experimental measurement chamber. The following parts are shown: (a) PCB shimming coil, (b) $\beta$-detectors, (c) \ce{^{1}H} magnetometer, (d) sample ladder, (e) $\beta$-windows, (f) rf-excitation coil, (g) collimator, (h) front viewport, (i) bottom plate with viewport, and feedthroughs, and two empty flanges for expansion purposes (j,k). See text for further details.}
    \label{fig:ExplodedChamber}
\end{figure}

As can be seen in the exploded view of the experimental chamber in Figure \ref{fig:ExplodedChamber}, a pair of double-layer plastic scintillator detectors were placed close to both magnet poles, i.e. along and opposite to the direction of spin polarisation.
Within the detectors, both 2-mm scintillator layers (Eljen Technology EJ-212) were insulated against light travelling between them with a 23.62~$\mu$m layer of aluminum.
Silicon Photon Multipliers (SiPMs) (onsemi MICROFJ-60035) were coupled to the scintillators, and the signal was amplified by on-board operational amplifiers.
This kept the detector system small, power efficient, and very resistant to external electronic noise sources.
Because of the use of SiPMs, the detectors were not affected by the 1.2T magnetic field, thus eliminating the need for light guides. 

\subsection{Sample exchange system}
\begin{figure}
    \centering
    \includegraphics[height=1.2\linewidth]{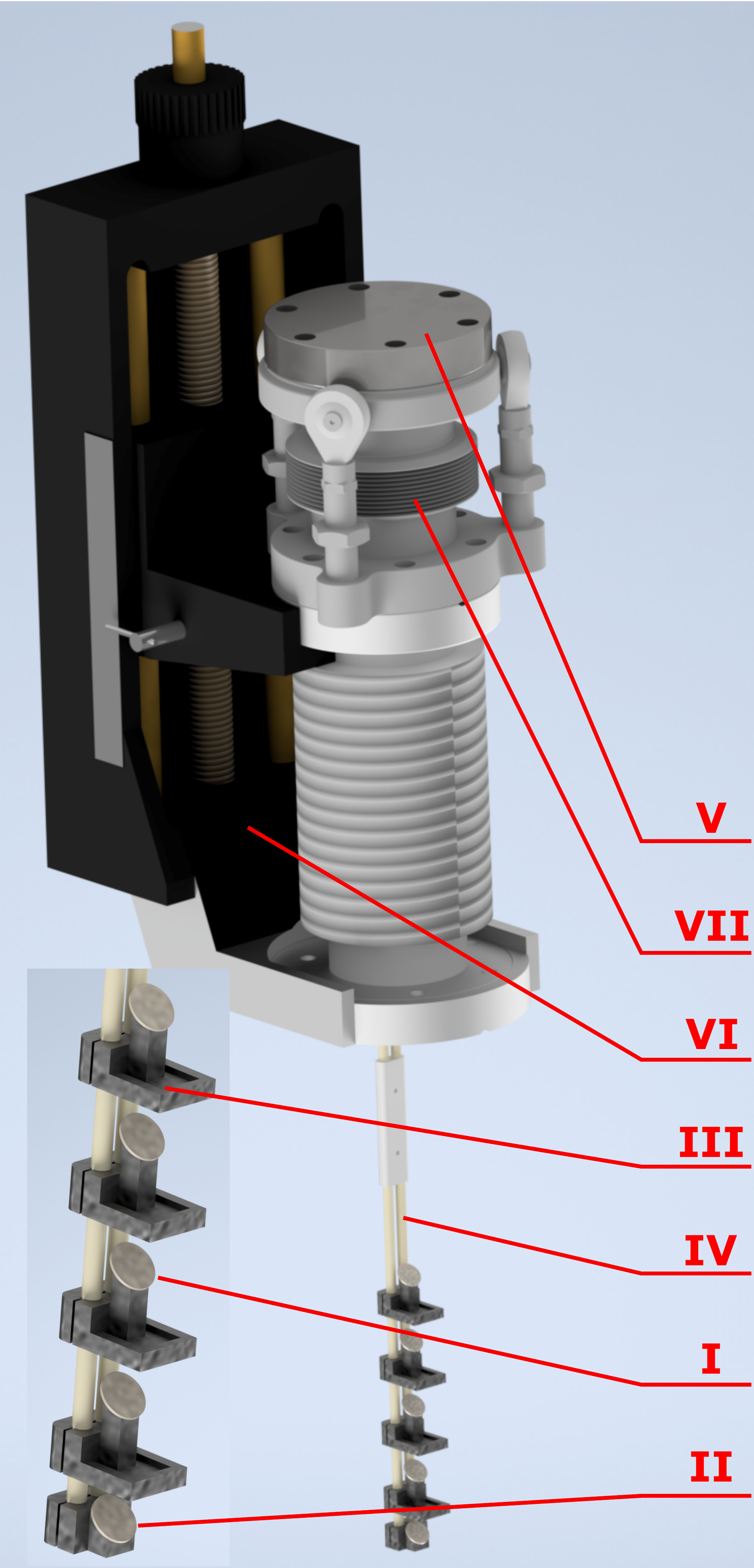}
    \caption{Sample exchange ladder that allows for the rapid exchange of samples. Shown are (i) the mica sheet sample holder, (ii) reference crystals, (iii) PLA mounting structures, (iv) ceramic rods, (v) vacuum flange, (vi) the linear translation stage and (vii) tilt adjust.}
    \label{fig:SampleLadder}
\end{figure}

To reduce the time spent on changing liquid samples, a sample ladder shown in Figure \ref{fig:SampleLadder} was developed.
The entire ladder fits within the 30~mm internal diameter of the rf excitation coil.
Individual sample holders, made of 8~mm diameter mica sheets (i) or reference crystals (ii), are mounted on 3D printed Polylactic acid (PLA) structures, (iii) connected to two ceramic rods (iv), which are fixed to a vacuum flange (v).
This flange, in combination with a linear translation stage (vi) and tilt adjust (vii), allows for coarse and fine adjustments of the sample position in all 3 dimensions.

To align the sample a theodolite is used to look through the end of the beamline at the back of the sample ladder.
The centre of the mica is indicated by a contrasting dot made at the back of the PLA structures.
This dot is used to position the sample in the centre of the pole axis and the vertical axis by changing the position of the linear stage and tilt adjust.
To centre the sample in the beam axis a similar dot on the lowest sample holder is observed through a viewport on the bottom-plate of the experimental chamber (Figure \ref{fig:ExplodedChamber} (i)) as the tilt adjust is changed.
After the tilt adjust has been correctly set only the position of the linear stage had to be moved to change between the different sample positions.

The sample exchange system allows to measure up to five samples in rapid succession, reducing the time spent changing samples (more than an hour per sample) during the measurement time by a factor of five.

\subsection{Experimental vacuum chamber}
The experimental vacuum chamber, shown in Figure \ref{fig:ExplodedChamber}, was designed and CNC-machined to tightly fit in-between the magnet poles, and to maximize the available space inside it.
On each side of the vacuum chamber, a plate housing a Printed Circuit Board (PCB) shimming coil on one side (a), and a $\beta$-detector on the other (b), is mounted on the outside of the chamber.
Concentric with the magnet, 200~$\mu$m thick aluminum sheets act as "$\beta$-windows" (e).
They form the boundary between the vacuum inside the chamber and the atmospheric pressure outside, while still allowing the $\beta$-particles to pass through to the detectors with as little absorption as possible.
Inside, the vacuum chamber (100x40x200 mm) accommodates a 30~mm inner-diameter rf-excitation coil (f), the \ce{^{1}H} magnetometer (c), and movable sample ladder (d), with a NaF reference crystal and mica sample holders.
At the front of the chamber, there is an in-vacuum pocket that can be used to mount a collimator (g) or a differential pumping cone.
The upper part of the front side houses a large viewport for sample observation (h). 
The bottom of the chamber is closed off with a bottom plate equipped with a viewport for sample alignment purposes (i) and two vacuum feedthroughs with SMA connectors for rf connections.
In the back, two additional flanges (j,k) are available for future setup expansions, e.g. sample manipulation and temperature control.
\section{Results}
In this part, we present the results achieved thanks to the upgrades described in the previous section. 

\begin{figure}
    \centering
    \includegraphics[width=\columnwidth]{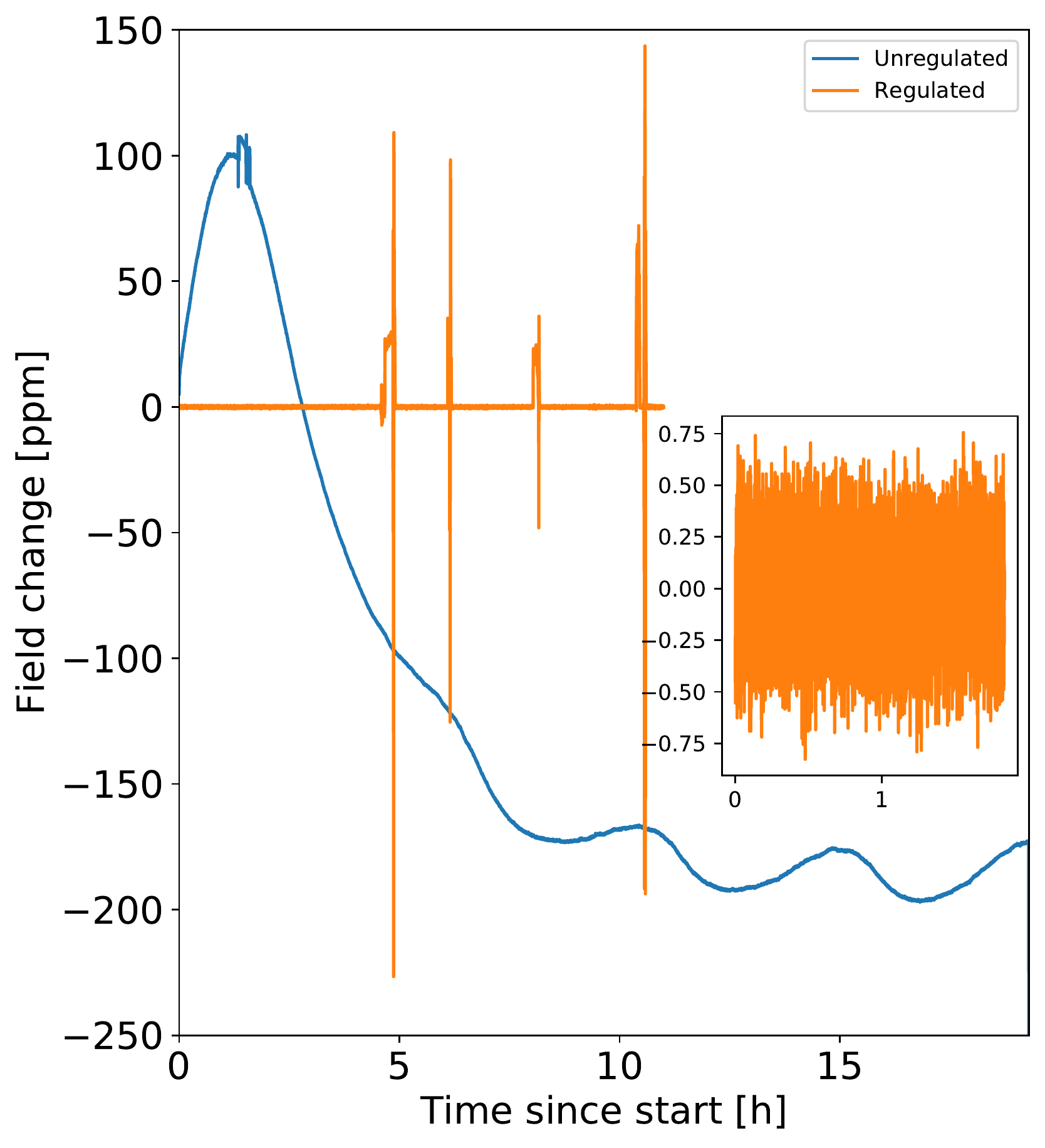}
    \caption{The change of the magnetic field compared to a reference value over time with the locking system off (blue) and on (orange). The insert shows a zoom of the regulated data. The spikes in the data are due to interventions on the experimental chamber requiring the regulation to be temporarily turned off. Once turned back on, the system cycles through most of its range before regaining the locked frequency. The initial drift of the unregulated data is due to slow equilibration of the magnet and power supply. The subsequent oscillations are caused by fluctuations in the cooling water and ambient temperature.}
    \label{fig:FieldLock}
\end{figure}

The implemented locking system has allowed us to effectively stabilize the magnetic field down to the ppm level.
As shown in Figure \ref{fig:FieldLock}, the fluctuations of the magnetic field were substantial when the locking system was turned off. 

When powered without the locking system, the magnet exhibited significant field instability while the power supply and magnet coils reached a temperature equilibrium.
This can take up to 10~hours.
The field change observed during this period has been as high as 300 ppm.
Following this initial instability the field shows an oscillation with a period of roughly 5~hrs and an amplitude of 17~ppm, which correlates with periodic changes in the ISOLDE cooling water temperature.
By comparison, with the locking system enabled (see the insert of Figure \ref{fig:FieldLock}), the field deviates around a set point by approximately $\pm$ 0.6~ppm.
The disturbances in the otherwise smooth lines are caused by manual interventions in the direct vicinity of the magnet during which the regulation system was turned off.
Upon restarting the system it cycles through most of its range before finding back the locked frequency.
This is the cause of the large spikes at the end of the intervention.

\begin{figure}
    \centering
    \includegraphics[width=\columnwidth]{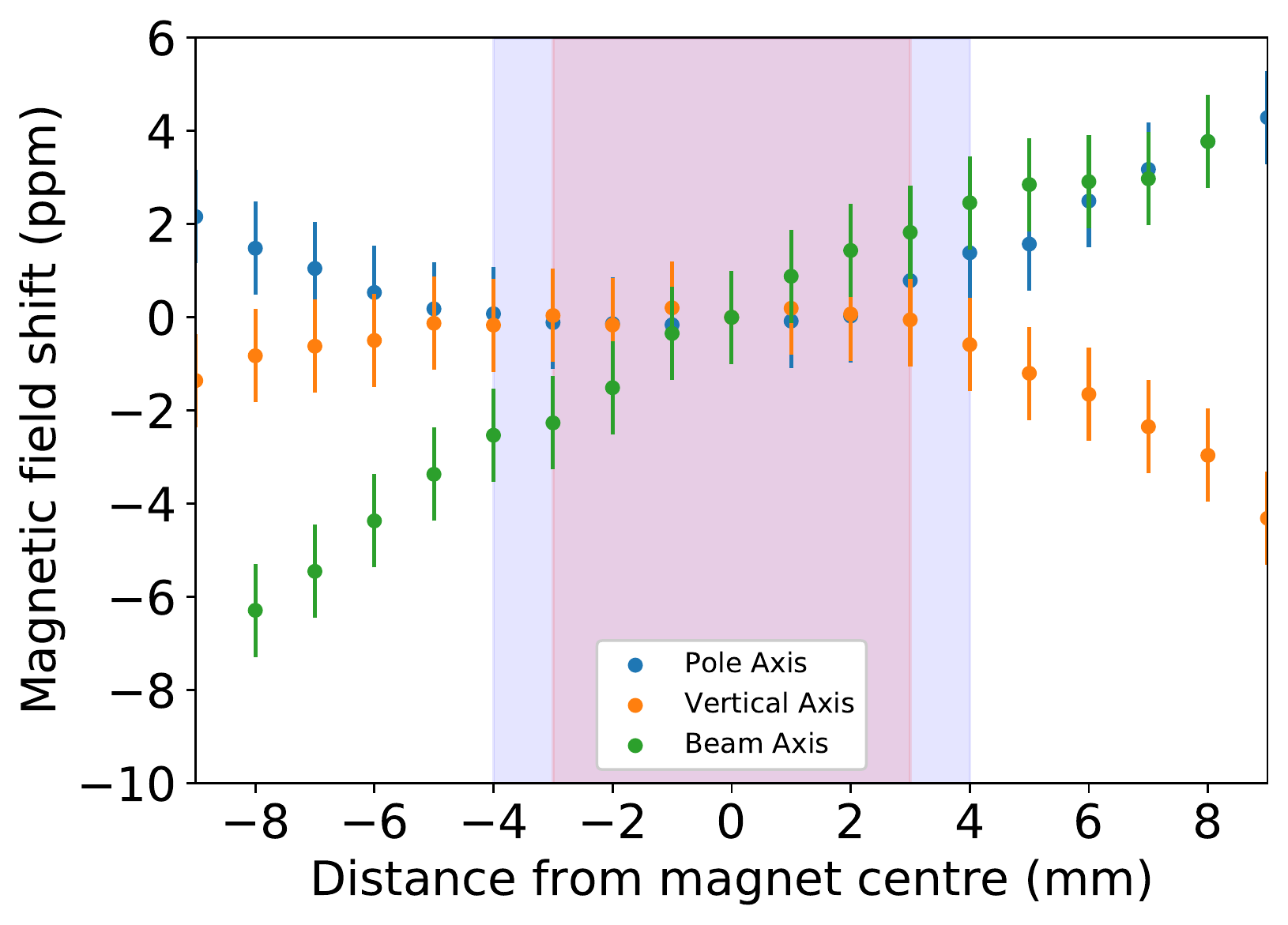}
    \caption{Final shimming results for the magnetic field along all three axes, relative to the magnetic field in the centre of the magnet (taken as 0 ppm). Blue symbols:  field deviation along the central pole-to-pole axis, orange symbols: deviation along the vertical axis, green symbols: deviation in the horizontal (beam) direction. The shaded blue area indicates the sample space in the pole-to-pole direction, and the shaded red area the sample space in the other two directions.}
    \label{fig:ShimCompare2}
\end{figure}

The shimming coils improve the homogeneity of $B_0$ within the sample volume to the single digit ppm level in every direction as shown in Figure \ref{fig:ShimCompare2}. With the shimming coils turned off, $B_0$ follows a parabola deviating by over 20~ppm from the center to 9~mm outwards on either side along the pole-axis.
With the shimming coils turned on, the deviation is reduced to only 4~ppm over the same distance.

\begin{figure}
    \centering
    \includegraphics[width=\columnwidth]{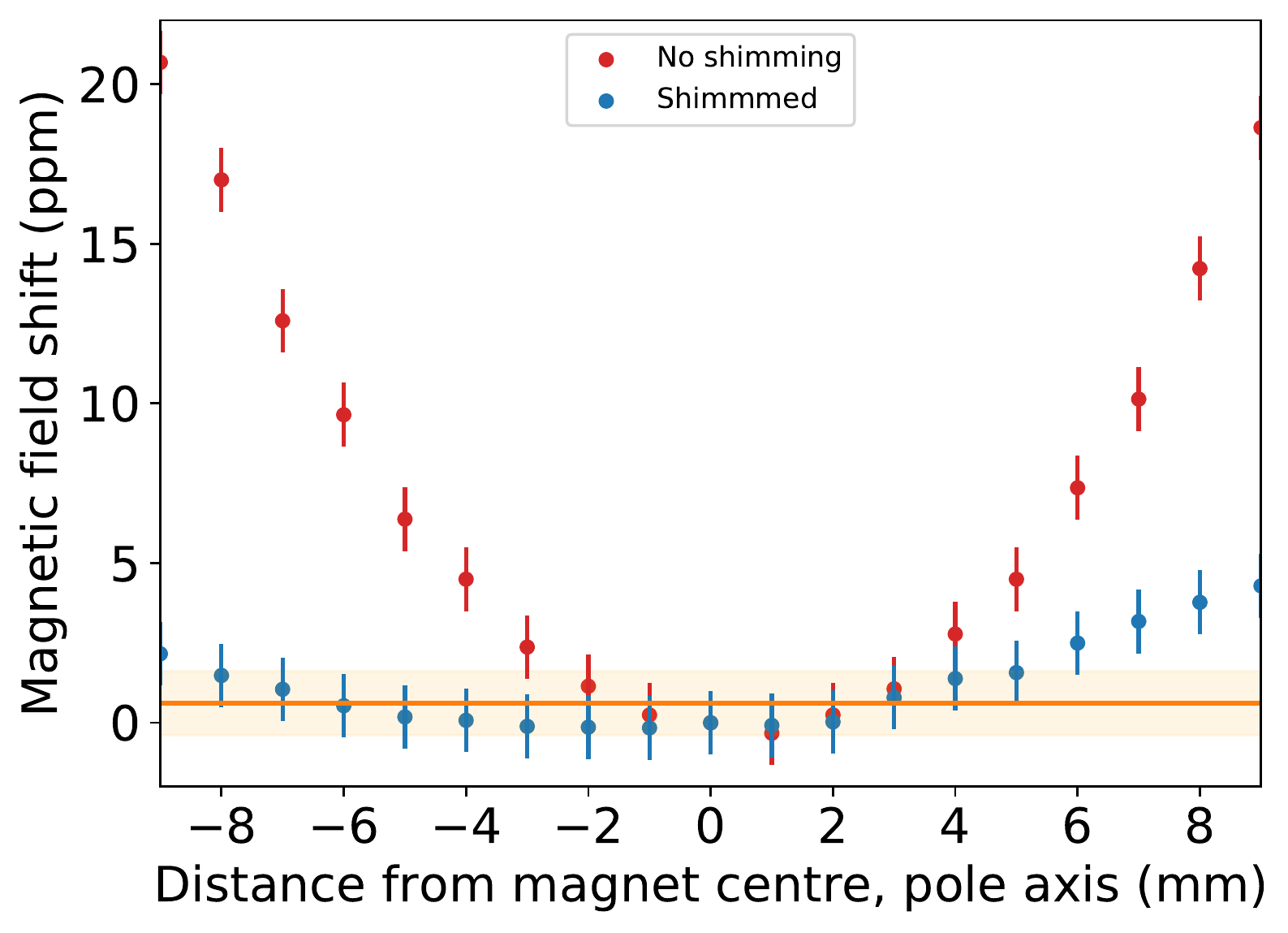}
    \caption{Spatial deviation of the magnetic field when going away from the magnet's center towards its poles. Red symbols: shimming coils turned off; Blue symbols: shimming coils on; Orange line shows the median field across the sample site and the shade area indicates a 1~ppm difference from this line.}
    \label{fig:ShimCompare}
\end{figure}

Because the sample is oriented with a 45 degrees angle to the horizontal plane (Figure \ref{fig:SampleLadder}) the sample volume extends 3~mm from the magnet center in the vertical and beam axis directions, and 4mm in the magnet-pole axis. Figure \ref{fig:ShimCompare2} shows the final shimmed magnetic field ($B_0$) along all three axes around the centre of the magnet.
This is the highest achieved homogeneity across the three combined dimensions, as trying to optimize the homogeneity in one axis further, decreases it in the others dimensions.
Between the magnet poles and the vertical axis $B_0$ follows a flat parabola with the center approximately in the magnet's middle, resulting in a maximum deviation from the median field along these axes at the sample site of 0.8 and 0.2~ppm respectively.
However, along the beam axis $B_0$ follows almost a straight line, resulting in a maximum deviation from the median field of 2.1~ppm.

\begin{figure}
    \centering
    \includegraphics[width=\columnwidth]{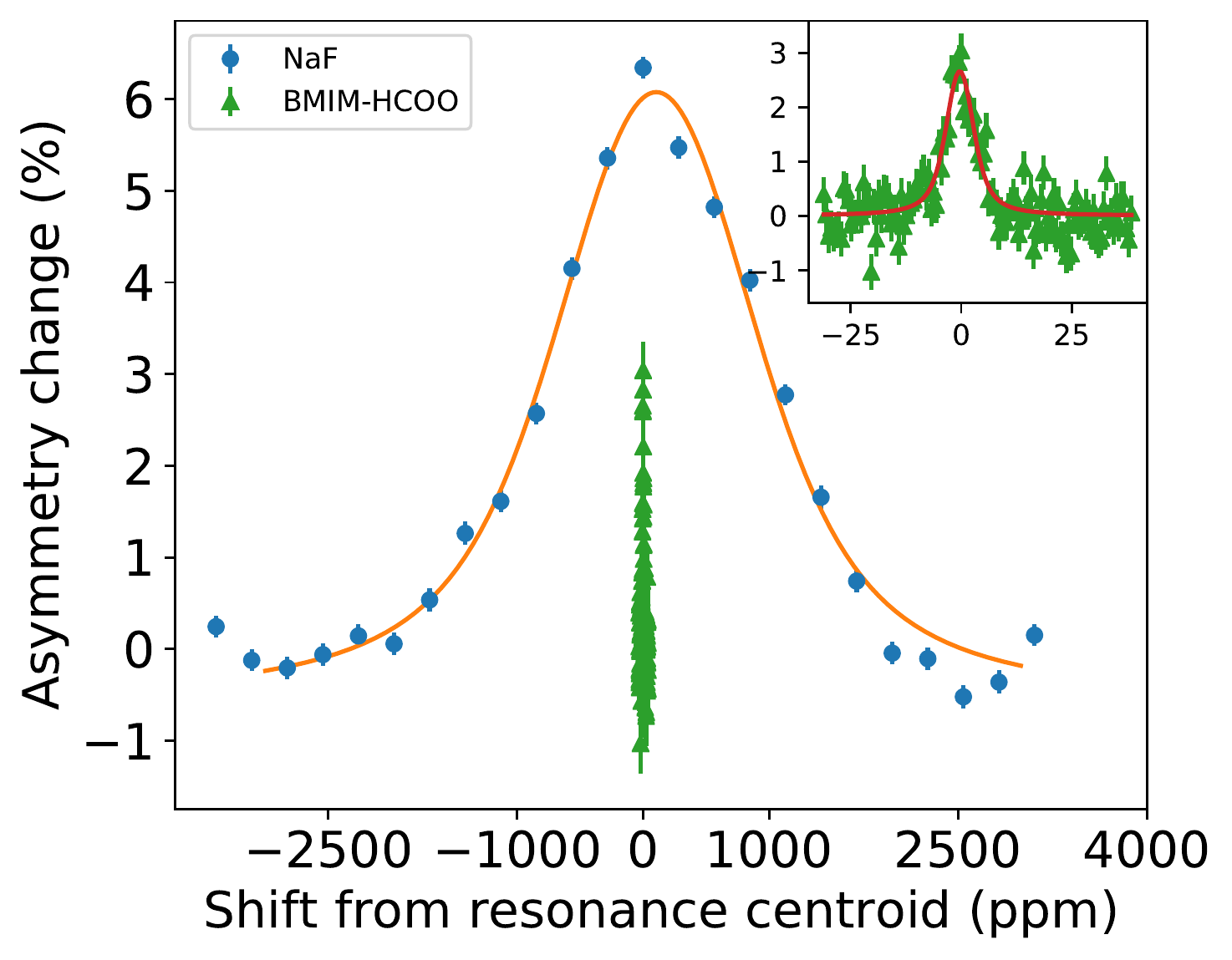}
    \caption{A comparison between the resonance width of \ce{^{26}Na} in a solid state host (NaF; blue), and a liquid state host (BMIM-HCOO; green). The deviation from the centre of each peak is given on the x-axis. The insert shows the resonance in the liquid host.}
    \label{fig:SolidVsLiquid}
\end{figure}

%Solid-State vs. liquid state beta-NMR
This field stability has allowed us to determine the magnetic moment of an unstable nucleus with two orders of magnitude higher accuracy than possible before.
Until recently\cite{Harding2020MagneticBiology, Mihara2019}, studies of the magnetic moment of exotic nuclei through $\beta$-NMR have relied on solid-state hosts as implantation sites \cite{Yordanov2007,Rogers1986, Connor1962, Matsuta1995}.
In the case of sodium, magnetic moment studies were performed using a NaF crystal, with cubic lattice structure, which retained the nuclear spin-polarisation for up to 40~s \cite{Kowalska2017, Keim2000}.
Resonances were achieved with a full width at half maximum (FWHM) in the order of $10^{-4}$ of the resonance frequency \cite{Keim2000}.
However, with liquid-state hosts, it is possible to obtain resonances with a FWHM down to two orders of magnitude smaller, whilst retaining the nuclear polarisation long enough to observe it through $\beta$-NMR.
The narrowing of the resonance lines is due to fast molecular tumbling within the liquid sample, effectively averaging out the anisotropic contributions found in solid-state NMR spectra.
Figure \ref{fig:SolidVsLiquid} shows a comparison the resonance width of \ce{^{26}Na} in a solid state host (NaF; blue), versus a liquid state host (BMIM-HCOO; green).
The insert in Figure \ref{fig:SolidVsLiquid}, is a zoomed in part of the spectra in order to show the resonance from the liquid host.

\section{Discussion}
In this article, we have presented in detail the technical solutions that allow to perform $\beta$-NMR studies in liquid samples with high absolute accuracy.

One of the key innovations we introduced is the magnetic field locking and measurement system that maintains the stability of the magnetic field down to the ppm level over hours and which allows us to determine the absolute value of the field with ppm accuracy.
Another improvement is the PCB-shimming coils, which improve the magnetic-field homogeneity to 1-5~ppm within the volume of the sample.
Furthermore, the use of state-of-the-art SiPMs has allowed us to obtain a very compact $\beta$-detection system opening up the space to observe the behavior of liquid samples in real time.
Finally, a special sample handling system shortened the time spent on changing samples by a factor of 5.
With the combination of these features, frequencies of $\beta$-NMR resonances in different hosts have been measured with unprecedented accuracy, paving the way for new applications of the technique.
At the same time, further improvements are possible, and these and some general characteristics are covered in the following sections.

\subsection{Locking system}
The present field locking system reads out the magnetic field every 200~ms, and adjusts the current at least once per second.
This measurement rate is 5 times faster than the previously used commercial Metrolab PT2025 magnetometer (Geneva, CH).
In addition, the probe is smaller and is fully vacuum compatible.
However, the rate at which the current in the electromagnet can be adjusted is the limiting step in the feedback loop.
Specifically, the control and adjustment of the stepper-motor-driven variable resistor takes longer than the measurement time of the \ce{^{1}H} NMR magnetometer.
Decreasing the time it takes to adjust the current can improve the field control further by making use of every measured \ce{^{1}H} frequency, thus reducing the time interval between consecutive field adjustments.

The \ce{^{1}H} magnetometer's other strength comes from the fact that it is based on the \ce{^{1}H} NMR frequency in water.
The absolute NMR shielding for \ce{^{1}H} in water has been calculated much more accurately than in other liquid or solid hosts, thus leading to a more accurate magnetic field readout at the NMR probe's position.
The determination of the absolute magnetic field can be done concurrently with $\beta$-NMR measurements.
Relative measurements via the resonance frequency of a reference $\beta$-NMR nucleus in a known host, which is both time-consuming and less accurate, are therefore unnecessary. 

Furthermore, our \ce{^{1}H} magnetometer uses pulsed NMR, which eliminates the need to modulate the static magnetic field (as is done in the PT2025 probe).
Thanks to this, the magnetometer does not disturb the magnetic field experienced by the probe nuclei at the sample location.
Additionally the diameter of the magnetometer's rf coil is much smaller than the continuous-wave $\beta$-NMR excitation coil and the applied frequencies are far apart, excluding cross-talk between the two rf coils.

At the same time, because the $\beta$-NMR coil is relatively large, the \ce{^{1}H} probe is positioned 25~mm away from the sample location, where the field is different and less homogeneous than at the sample's location in the magnet's center.
Though this does not affect the locking capabilities of the system, it leads to a field offset that needs to be measured and corrected for, in order to accurately determine the magnetic field at the sample position.
As the field's gradient increases, so does the uncertainty of this correction.
Even a small change of the probe's position can then result in a slightly different measured field.
At present, this uncertainty is the largest contribution to the total uncertainty in the value of the magnetic field at the sample's position. 
Higher accuracy in the determination of the magnetic field at the sample position can be reached, e.g. by measuring the magnetic field very close to, or even at, the exact position of the sample.
For example, in conventional NMR spectrometers, this is done by measuring the resonance frequency of \ce{^{2}H} present in heavy water (\ce{D_{2}O}) mixed into the solvent.
In $\beta$-NMR, protons in the solvent molecules can be used for the same purpose if an excitation and pickup coil is used to acquire a conventional \ce{^{1}H} NMR signal from the sample itself.

\subsection{Shimming coils}
The main achievement of the shimming coils was to reach 1~ppm homogeneity over a distance of several millimeters.
Because the coils are made as PCB boards, they are very robust compared to wire wound coils.
They are also cheap and easy to manufacture reproducibly.
The optimal current setting can easily be calculated using Eq. \ref{eq:ShimCur}, requiring only minor fine tuning afterward.

A consequence of this design, which is dependent on the distance between the poles of the electromagnet, is that the corrections of the field in the axial and radial directions are coupled.
This means a compromise must be found between the field homogeneity in different directions.
Thus, a small gradient arising from the tail of the transitional field used to adiabatically rotate the atomic spin and consecutively decouple the nuclear and electron spin of the incoming beam of probe isotopes \cite{Kowalska2017}, can not be corrected by this system.

The magnetic field homogeneity could be further improved by analysing the field map, and by designing additional coils to correct for remaining field inhomogeneities.
This would however quickly become very complicated, as one would need to take into account not only the field at the sample site, but also along the beam path, to make sure that field adjustments made by any additional shimming coils do not negatively affect the aforementioned transitional field.
The extra space required for additional shimming coils will also be a limiting factor. 

\subsection{Sample handling system}
Because the sample handling system reduces the time spent on exchanging samples 5 fold, it significantly increases the number of samples that can be tested during an experimental campaign.
It also allows for the fine adjustment of each sample's position, resulting in reproducible sample alignment.
When used in combination with the upstream collimator at the front of the experimental chamber, the sample handling system allows to have the polarised beam implanted solely in the well-aligned sample.
This prevents background $\beta$-counts that could arise from parts of the beam being implanted in materials other than the sample.

During previous experiments, a NaF crystal was mounted at the bottom of the ladder, for reference and inspection purposes.
The mica substrate used to hold the liquid samples is compatible with biological samples, easy to clean, and allows to create a thin, sub-millimeter layer of the liquid hosts (corresponding to 5-20~$\mu$L of liquid), which also makes it possible to use solvents only available in very small quantities. 
As with all other parts situated between, or close to the magnet poles, the sample handling system is made of non-magnetic material, and thus does not disturb the static magnetic or the rf fields.

The sample handling system is actuated manually.
A clear improvement from the operational standpoint would be the automation of sample positioning.
In addition, it does not yet include sample temperature measurements and control. 
These will be implemented in future developments of the system.
Due to the lack of an efficient vacuum-liquid barrier and the sample orientation at 45° to the vertical, the sample handling system is only suitable for relatively viscous vacuum-resistant solvents.
Low viscous solvents would simply flow off of the sample holder and non-vacuum-resistant solvent will evaporate and/or freeze due to the low pressure.

Ideally, for biochemical studies, one would want to perform measurements in water, as this is what is routinely done in conventional NMR studies.
To achieve this goal, new technical approaches will be necessary to combine the no/low-vacuum environment required by water (because of its high vapour pressure), with the high-vacuum environment of RIB facilities.
Solutions that warrant investigation for that specific purpose are differential pumping setups, allowing poor vacuum around the sample, combined with better vacuum along the path of the beam of unstable nuclei, or the encapsulation of samples to protect the high vapor pressure solvents from evaporation in the vacuum.
Both of these approaches present their own hurdles.
In the case of a high final pressure in a differential pumping setup, the gas present right before the sample will significantly scatter the incoming beam due to beam-gas interactions, which will significantly decrease the $\beta$-NMR SNR.
In the case of encapsulation, the window material must be able to withstand the pressure gradient, but also transmit most of the beam without affecting its spin polarisation.
Interactions between the beam and the window heavily depend on the material used for the window, its thickness, and the energy of the beam.
For low energy beams such as in our setup (up to 50~keV), the window should be less than 100~nm thick, even possibly only 50~nm thick, which makes it very brittle.
For high energy beams used at in-flight facilities, the window can be thicker.
This allowed Mihara \textit{et al.} \cite{Mihara2019}, to perform a $\beta$-NMR measurement of \ce{^{17}{N}} in water.

\subsection{Beta detectors}
The current beta-detection system is a significant improvement compared to previous systems, which also relied on two pairs of thin plastic scintillators, however these were coupled to conventional PMs.
Since conventional PMs are very sensitive to magnetic fields, they needed to be covered in Mu-metal and kept almost 1~m away from the center of the magnet, necessitating the use of light guides.
The whole system took up lot of space, making any manipulation impractical, and mostly preventing visual inspection of the implantation site.
It was also very heavy, and thus cumbersome to assemble and to get ready for operation.
The present system is lighter and more compact, with $\beta$-detectors (scintillator and SiPMs) located inside the magnet, but still outside of the NMR chamber.
In the future it should be possible to craft a set of detectors to be placed closer to the sample in vacuum.
These could be made significantly smaller if the covered solid angle is kept constant.
Besides this, one can consider designs with more complex geometries, so as to take full advantage of the beta decay angular distribution and the effect of the magnetic field on beta-particle trajectories. 

\subsection{Experimental chamber}
Our experimental chamber design was to a large degree determined by the vacuum regime (no need for vacuum better than $10^{-6}$~mbar), the magnet to be used, and by respecting previously made design decisions.
It allows to accommodate all necessary elements: $\beta$-NMR excitation coil, \ce{^{1}H} magnetometer, samples, detectors and shimming coils.
Additionally, it has 2 viewports, combined with the viewport at the end of the beamline, allowing to check the alignment, circular polarisation, and power of the laser beam, to define the position of each sample holder after sample change and finally to observe the behaviour of the liquid sample on the substrate.
The chamber is made of heat treated (T6) 6082 aluminium alloy and has standard-sized flanges compatible with the UHV standard CF (using elastomer instead of copper gaskets), which makes the installation and alignment easier. 

The rf-excitation coil was designed to not obstructed the incoming beam, allow for the maximum solid angle of $\beta$-radiation through the "$\beta$-windows, and the best homogeneity of the applied rf-field perpendicular to B$_{0}$ over the sample volume given these two boundary conditions. 

Different magnet configurations, however, come with significantly different constraints.
For example, for a solenoidal magnet, the chamber design will not easily give access to a viewport for sample observation. The design of a sample ladder would also not be a trivial task.

%%Implications: Generalizations, Significance, Recommendations.
\section{Conclusion}%\subsection{General remarks}
The upgrades described in the present manuscript open new avenues for $\beta$-NMR studies.
They have already allowed for the determination of the magnetic moment of the short lived nucleus \ce{^{26}Na} with ppm precision \cite{Harding2020MagneticBiology}. 
This achievement, together with accurate hyperfine structure measurements \cite{Stroke2000}, should allow for the study of the distribution of nuclear magnetism through the hyperfine anomaly (the effect of the non-point like size of the nucleus on the hyperfine structure\cite{Persson2013}).
The distribution of nuclear magnetisation can in turn shed light on open questions in nuclear structure physics, in particular about the distribution of neutrons inside the nucleus \cite{Hagen2016,Thiel2019}.

Another application that these upgrades will be essential for, is the study of interactions between metal ions and biomolecules with $\beta$-NMR \cite{Crichton2019BioInorgChem,Crichton2020PractBioInorgChem}.
One example of such interactions is the interaction between DNA G-Quadruplexes and sodium or potassium ions \cite{Carvalho2020,Wu2003,WONG2005,wong2009}.
Presently, we are exploring the most suitable potassium probes for G-quadruplexes studies \cite{Karg2020proposal}, as well as isotopes of several other elements relevant to protein folding \cite{Jansco2017}.
Together with our previous $^{26}$Na measurement \cite{Harding2020MagneticBiology} this will be of direct use in our current investigations into this topic.
\section{Acknowledgements}
This work was supported by the European Research Council (Starting Grant 640465),
the UK Science and Technology Facilities Council (ST/P004423/1),
EU project ENSAR2 (654002),
the Ministry of Education of the Czech Republic (LM2018104),
the Wolfgang Gentner Programme of the German Federal Ministry of Education and Research (05E15CHA),
and by the Swiss Government Excellence Scholarships for Foreign Scholars program.

\bibliographystyle{Template/model1-num-names}
\bibliography{References}

\end{document}